\documentstyle[prd,aps,twocolumn,epsf,floats,amsfonts,amssymb,amsmath]{revtex}
\newcommand{\mlab}[1]{\label{#1}}
\newcommand{\EX}{{$\eta$-$\xi$ }}

\newcommand{\etad}{\eta_{_\delta}}
\newcommand{\xid}{\xi_{_\delta}}
\newcommand{\I}{{_I}}
\newcommand{\II}{{_{I\!I}}}
\newcommand{\III}{{_{I\!I\!I}}}
\newcommand{\IV}{{_{I\!V}}}
\newcommand{\IId}{{_{I\!I_{\!\delta}}}}

\newcommand{\IQ}{{_I,_{I\!I}}}
\newcommand{\R}{{\Bbb R}}
\newcommand{\Vk}{{\bf k}}
\newcommand{\Vp}{{\bf p}}
\newcommand{\Vx}{{\bf x}}

\begin{document}
\twocolumn[\hsize\textwidth\columnwidth\hsize\csname@twocolumnfalse\endcsname
\preprint{Imperial/TP/97-98/79, hep-th/98}

\title{Geometric Background for Thermal Field Theories}

\author{Massimo Blasone${}^{\flat\natural}$,
Yuan-Xing Gui${}^{\flat\sharp}$
and Francis Vendrell${}^{\flat\dagger}$ \vspace{2mm}}

\address{${}^{\flat}$Blackett Laboratory, Imperial College, Prince
Consort Road, London SW7 2BZ, United Kingdom}

\address{${}^{\natural}$Dipartimento di Fisica, Universit\`a di
Salerno and INFN, I-84100 Salerno, Italy}

\address{${}^{\sharp}$Department of Physics, Dalian University of
Technology, Dalian 116024, China}

\address{${}^{\dagger}$D\'epartement de physique th\'eorique, Ecole de
physique, Universit\'e de Gen\`eve, CH-1211 Gen\`eve 4, Switzerland}

\date{\today}

\maketitle

\begin{abstract}

We study a new spacetime which is shown to be the general geometrical
background for Thermal Field Theories at equilibrium. The different
formalisms of Thermal Field Theory are unified in a simple way in
this spacetime. The set of time-paths used in the Path Ordered
Method is interpreted in geometrical terms.
\end{abstract}

\vspace{8mm} ]



\section{Introduction}

Thermal Field Theory (TFT) \cite{DAS,LEB,LW87} is the generic name
denoting Quantum Field Theories (QFT) at finite temperature, such as
the Matsubara or Imaginary Time (IT) formalism \cite{DAS,LEB,LW87},
Thermo Field Dynamics (TFD) \cite{DAS,UM0}, and the Path Ordered
Method (POM), which comprises the Closed Time Path (CTP) formalism
\cite{DAS,LEB,LW87} as a particular case. These distinct approaches
result from different efforts to introduce a temperature within the
framework of quantum field theory.

\medskip
The IT formalism exploits the analogy of the temperature with
imaginary time in calculating the partition function, and within this
formalism the two-point Green's function is given by the Matsubara
propagator. In contrast, the POM formalism and TFD both deal with real
time. These two last approaches lead to the same matrix structure for
the propagator in equilibrium, although their founding ideas are
different.  In the POM formalism, the temperature is introduced by
adding a pure imaginary number to the real time and by choosing a
special path in the complex-time plane. The propagators are calculated
by taking functional derivatives of a path integral over fields, the
so-called finite-temperature generating functional. On the other hand, 
in TFD, the field algebra is doubled and the temperature is contained
explicitly in the resulting ``vacuum'' state, which is then referred
to as the ``thermal vacuum''.  The propagators are expressed in this
last formalism as expectation values of time-ordered products of
quantum fields with respect to the thermal vacuum.

\medskip
An important and interesting question is whether such different
formalisms have some roots in common and if their features can be
understood in a deeper way so that they appear unified. A clue to an
answer may be found in the well-known discovery of Hawking \cite{HAW}
that temperature arises in a quantum theory as a result of a
non-trivial background endowed with event-horizon(s), in this case a
black-hole spacetime. Rindler spacetime, the spacetime of an
accelerated observer, was also shown later to exhibit thermal features
\cite{Rin,BIDA}.  There is also a flat background with a non-trivial
structure which exhibits thermal features, the so-called \EX spacetime
\cite{GUI90,GUI92,GUI92ferm}, which has besides revealed itself useful
for discussing properties of black holes \cite{FV}.  We believe that
\EX spacetime is the generic example of a spacetime exhibiting thermal
properties in the sense that fields and states in this spacetime look
everywhere as if they were immersed in a thermal bath contained in a
Minkowski background \cite{GUI90,GUI92}.

\medskip
\EX spacetime is a flat complex
manifold with topology $S^1\times R^3$. Its Lorentzian section
is made up of four copies of Minkowski spacetime glued together along
some of their past or future null hyperplanes infinities. In
Kruskal-like coordinates, which cover the entire \EX spacetime, 
the metric is singular on these hyperplanes. For
this reason they are {\em formally} called ``event-horizons''.
Their existence leads to the doubling of the degrees of freedom of the
fields. The vacuum propagator on the Lorentzian section corresponds to
the real-time thermal matrix propagator. In the Euclidean section of
Kruskal coordinates, the time coordinate is periodic, so that the
propagator is equal to the Matsubara propagator there.

\begin{figure}[t]
\centerline{\epsfysize=2.0truein\epsfbox{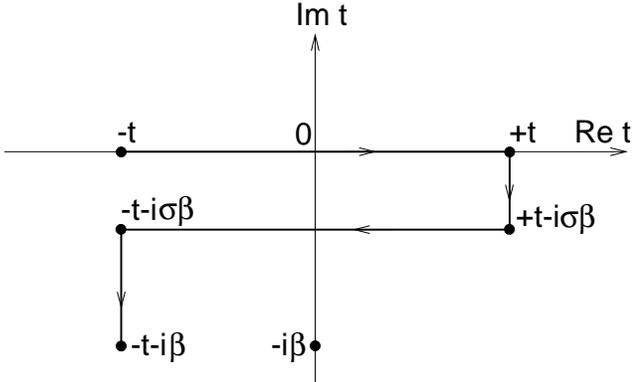}}
\caption{The time path used in POM. The parameter $\sigma$ ranges from
the value $\sigma=0$ (Closed Time Path) to $\sigma=1$.}
\label{fig1}
\end{figure}

\medskip
In real-time TFTs, there is a freedom in the parameterization of the
thermal matrix propagator. In the POM formalism, this parameterization
is related to the choice of the path in the complex-time plane going
from $t=0$ to $t=-i\beta$ which is not unique \cite{NS} (see
Fig.~\ref{fig1}).  In TFD, different parameterizations of the
Bogoliubov thermal matrix are possible.  It is important to stress
that although the choice of the parameterization is irrelevant in the
thermal equilibrium case since it does not affect physical quantities
in real-time TFTs, it nevertheless plays a role in the non-equilibrium
case, where the choice of a closed time path is the only useful one
\cite{H95,HGM92,EHUY92}.

\medskip
In this paper we show the relevance of other complex sections of \EX
spacetime than the Lorentzian and the Euclidean ones.  All possible
time paths used in the POM formalism, including the CTP formalism, and
the different parameterizations of the Bogoliubov matrices of TFD are
interpreted at a purely geometric level in these complex sections. The
geometric picture for TFTs is consequently enlarged.  We also show how
the different formalisms of TFT unify themselves in a very natural way
in this framework.  The generalization introduced here could be useful
in order to extend the geometrical picture of \EX spacetime to systems
out of thermal equilibrium.  We consider in detail the case of a
scalar field, the generalization to fermions being straightforward as
shown in Ref.~\cite{GUI92ferm}.

\medskip
The paper is organized as follows: In Section II, we define the \EX
spacetime in general and then we study its Euclidean and Lorentzian
sections, and their complex extensions.  In Section III, we
consider explicitly a quantum boson field in \EX spacetime. Section IV
is devoted to the study of the relationships between \EX spacetime and
TFTs, and to the unification of TFTs in the framework of \EX
spacetime.  Section V contains a discussion of some other features of
the extended Lorentzian section. Finally, Section VI is devoted to
conclusions.

\section{\EX spacetime}


\EX spacetime \cite{GUI90,GUI92} is a four-dimensional complex
manifold defined by the line element
\begin{equation} \mlab{guimetric}
ds^2 = \frac{-d\eta^2+ d\xi^2}{\alpha^2\left(\xi^2-
\eta^2\right)} + dy^2 + dz^2,
\end{equation}
where $\alpha\equiv 2\pi/\beta$ is a real constant and
$(\eta,\xi,y,z)\in{\Bbb C}$.  We shall use the symbol
$\xi^\mu\equiv(\eta,\xi,y,z)$ to denote as a whole the set of \EX
coordinates, but for simplicity we shall actually drop the index
$\mu$ when no confusion with the space-like coordinate is possible.

\subsection{Euclidean section}

The Euclidean section of \EX spacetime is obtained by assuming that
$(\sigma, \xi, y,z)\in{\Bbb R}$ where $\eta\equiv i\sigma$. The metric
in this section is given from Eq.~(\ref{guimetric}) by
\begin{equation} \mlab{euclidmetric}
ds^2 = \frac{d\sigma^2+ d\xi^2}{\alpha^2\left(\sigma^2+\xi^2\right)} 
+ dy^2 + dz^2.
\end{equation}
By use of the transformation
\begin{eqnarray} \mlab{euclidTrans}
\left\{
\begin{array}{rcl}
\sigma & =&  (1/\alpha)\,
\exp\left(\alpha x\right) \sin\left(\alpha\tau\right), \\[2mm]
\xi &=&  (1/\alpha)\,
\exp\left(\alpha x\right) \cos\left(\alpha\tau\right),
\end{array}\right.
\end{eqnarray}
the metric becomes that of the cylindrical Euclidean flat spacetime,
\begin{eqnarray}
ds^2=d\tau^2+dx^2+dy^2+dz^2,
\end{eqnarray}
where the time $\tau$ has a periodic structure, i.e.~$\tau\equiv\tau
+\beta$.

\subsection{Lorentzian section} \label{subsecAC}

\begin{figure}[t]
\centerline{\epsfysize=2.5truein\epsfbox{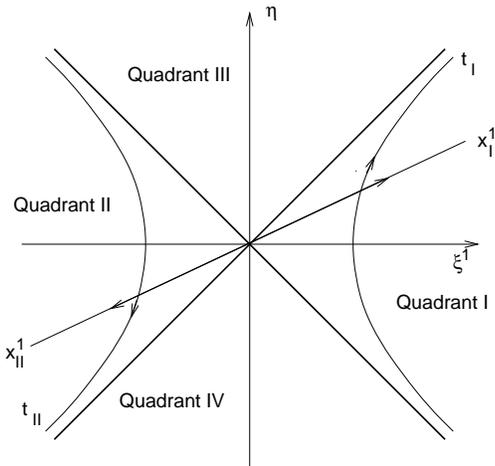}}
\caption{Lorentzian section of \EX spacetime: the solid lines
represent the singularities at $\xi^2-\eta^2=0$. On the straight lines
time is constant, while on the hyperbolas the Minkowski coordinate $x$
is constant. Note that time flows in opposite directions in regions
$I$ and $II$.}
\label{fig2}
\end{figure}

In the Lorentzian section the metric is given by Eq.~(\ref{guimetric})
where $(\eta, \xi, y,z)\in{\Bbb R}$.  This metric is singular on the
two hyperplanes $\eta=\pm\xi$ which shall be called the
``event-horizons''. They divide \EX spacetime into four regions
denoted by $R_\I, R_\II,R_\III$ and $R_\IV$ (see Fig.~\ref{fig2}).

\smallskip
In the first two regions, one defines two sets of tortoise-like
coordinates $x_\IQ\in\R^4$ by $x_\IQ=(t_\IQ,x_\IQ,y,z)$ where
\begin{eqnarray} 
&\mbox{in $R_\I$:} & \quad \left\{
\begin{array}{rcl}
\eta & =& +(1/\alpha)\,
\exp\left(\alpha x_\I\right)\sinh\left(\alpha t_\I\right), \\[2mm]
\xi &=& +(1/\alpha)\,
\exp\left(\alpha x_\I\right)\cosh\left(\alpha t_\I\right),
\end{array}\right.
\label{TransI}\\[3mm]
&\mbox{in $R_\II$:} & \quad \left\{
\begin{array}{rcl}
\eta & =& - (1/\alpha)\,
\exp\left(\alpha x_\II\right)\sinh\left(\alpha t_\II\right), \\[2mm]
\xi &=& - (1/\alpha)\,
\exp\left(\alpha x_\II\right) \cosh\left(\alpha t_\II\right).
\end{array}
\right. 
\label{TransII}
\end{eqnarray}
Similar transformations can be defined to cover $R_\III$ and $R_\IV$
(see Refs.~\cite{GUI90,FV}) but it shall be sufficient for our
purposes to consider only the first two regions.  The metric given in
Eq.~(\ref{guimetric}) becomes the Minkowski metric in these regions
and in the new coordinates,
\begin{eqnarray}
ds^2 = -dt^2_\IQ+dx_\IQ^2+dy^2+ dz^2.
\end{eqnarray}
Regions $I$ to $IV$ are copies of the Minkowski spacetime glued
together along the ``event-horizons'' making up the Lorentzian section
of \EX spacetime.

\smallskip
Although Eqs.~(\ref{TransI}) and (\ref{TransII}) are formally Rindler
transformations \cite{Rin}, the $(t_{\I,\II},x_{\I,\II},y,z)$
coordinates should not be confused with Rindler coordinates since in
those coordinates the metric is the Minkowski one. An observer whose
motion is described for example by
$(x_\I(t_\I),y_\I(t_\I),z(t_\I))=(x_0,y_0,z_0)$ is subjected to an
inertial motion, and thus cannot be identified with an accelerated
observer as in the Rindler case. In Eqs.~(\ref{TransI}) and
(\ref{TransII}), the role of the inertial and non-inertial coordinates
are actually reversed with respect to the Rindler case
\cite{Rin,GUI90}.

\bigskip
We now study the analytical properties of the transformations given in
Eqs.~(\ref{TransI}) and (\ref{TransII}). If $\xi^\pm=\eta\pm\xi$, we
rewrite them in the form
\begin{eqnarray}
&&\left\{\begin{array}{rcl}
\xi^+_>(t_\I,x_\I) &=&
+(1/\alpha)\exp\left[+\alpha\left(t_\I+x_\I\right)\right],
\\[2mm]
\xi^-_<(t_\I,x_\I) &=&
-(1/\alpha)\exp\left[-\alpha\left(t_\I-x_\I\right)\right],
\end{array}\right.
\\[2mm]
&&\left\{\begin{array}{rcl}
\xi^+_<(t_\II,x_\II) &=&
-(1/\alpha)\exp\left[+\alpha\left(t_\II+x_\II\right)\right],
\\[2mm]
\xi^-_>(t_\II,x_\II) &=&
+(1/\alpha)\exp\left[-\alpha\left(t_\II-x_\II\right)\right],
\end{array}\right.
\end{eqnarray}
where we have added the subscripts $<$ and $>$ to the variables
$\xi^\pm$ to indicates their ranges, i.e.~one has $\xi^\pm_>>0$ and
$\xi^\pm_<<0$. The reciprocals of these transformations are
\begin{eqnarray}
\left\{\begin{array}{rcl}
t_\I(\xi^+_>,\xi^-_<) &=& \displaystyle
\frac{1}{2\alpha}\ln\left(\displaystyle-\frac{\xi^+_>}{\xi^-_<}\right),
\\[4mm]
x_\I(\xi^+_>,\xi^-_<) &=& \displaystyle
\frac{1}{2\alpha}\ln\left(-\alpha^2\xi^+_>\xi^-_<\right),
\end{array}\right.
\label{RecI} \\[2mm]
\left\{\begin{array}{rcl}
t_\II(\xi^+_<,\xi^-_>) &=& \displaystyle
\frac{1}{2\alpha}\ln\left(\displaystyle-\frac{\xi^+_<}{\xi^-_>}\right),
\\[4mm]
x_\II(\xi^+_<,\xi^-_>) &=& \displaystyle
\frac{1}{2\alpha}\ln\left(-\alpha^2\xi^+_<\xi^-_>\right).
\end{array}\right.
\label{RecII}
\end{eqnarray}
Equations (\ref{RecI}) and (\ref{RecII}) are defined in regions $R_\I$
and $R_\II$ respectively. We now would like to extend analytically
these expressions to obtain the functions $t_\I(\xi)$, $t_\II(\xi)$
and $x_\I(\xi)$, $x_\II(\xi)$ defined in $R_\I\cup R_\II$. This
amounts to extend these expressions from positive or negatives values
of $\xi^\pm$ to their negative or positive values respectively.  In
order to do this, we choose to analytically extend the expressions
above in the {\it lower} half-planes of both the $\xi^+$ and $\xi^-$
complex planes for reasons which will become clear below. In other
words we assume that $-\pi\leq\arg\xi^\pm<\pi$, or equivalently that
the cuts in the $\xi^\pm$ complex planes are given by $\R_-+i0^+$.

\smallskip
It is not possible to perform the analytic extensions with respect to
the two variables $\xi^\pm$ {\it at once}, otherwise an erroneous
result would be obtained. To fix the ideas, we choose to perform the
extension first in the $\xi^+$ variable and then in $\xi^-$ (the
choice of the opposite order gives the same result for our particular
purposes).  If $\xi^\pm_<$ is the analytic continuation of $\xi^\pm_>$
from negative to positive values, one has
\begin{align}
\ln\left(-\xi^\pm_<\right) &= \ln\left(+\xi^\pm_>\right)+i\pi,
\\[2mm]
\ln\left(+\xi^\pm_>\right) &= \ln\left(-\xi^\pm_>\right)-i\pi.
\end{align}
This implies
\begin{eqnarray}
\ln\left(-\frac{\xi^+_<}{\xi^-_>}\right) &=&
\ln\left(-\frac{\xi^+_>}{\xi^-_<}\right) + i2\pi,
\\[2mm]
\ln\left(-\alpha^2\xi^+_<\xi^-_>\right) &=&
\ln\left(-\alpha^2\xi^+_>\xi^-_<\right),
\end{eqnarray}
which means that these expressions are the analytic continuations of
each others. In consequence, by using the variable $\xi$ one obtains
in $R_\I\cup R_\II$,
\begin{eqnarray}
\left\{ \begin{array}{rcl}
t_\II(\xi) &=& t_\I(\xi) +i\,\beta/2, \\[2mm]
x_\II(\xi) &=& x_\I(\xi).
\end{array} \right.
\label{ACtx}
\end{eqnarray}

\subsection{Extended Lorentzian section} \label{sub:ELS}

We now consider a class of complex sections of \EX spacetime obtained
from the Lorentzian section by shifting the Minkowski time coordinates
in the imaginary direction {\it only} in region $R_\II$,
\begin{eqnarray}\mlab{shift}
\begin{array}{lll}
&\mbox{in $R_\I\cup R_\III\cup R_\IV$:} 
\qquad & t_q \rightarrow {t_q}_{_\delta}=t_q,
\\[2mm] &\mbox{in $R_\II$:} \qquad & t_\II
\rightarrow t_{\IId}= t_\II + i\beta\delta,
\end{array} 
\end{eqnarray}
where $\delta\in[-1/2,1/2]$. The set of these sections shall be 
called ``extended
Lorentzian section.'' We denote by $R_\IId$ the image of the region
$R_\II$ through this shift. In $R_\IId$ the \EX coordinates become
complex variables and are transformed according to $(\eta,\xi)
\rightarrow (\etad,\xid)$ where, from Eq.~(\ref{shift}),
\begin{eqnarray}\mlab{mex2a}
\left\{
\begin{array}{rcl}
\etad &=& -(1/\alpha)\,\exp\left(\alpha x_\II\right)
\sinh\left[\alpha\left(t_\II+i\beta\delta\right)\right],
\\[2mm]
\xid &=&  -(1/\alpha)\,\exp\left(\alpha x_\II\right)
\cosh\left[\alpha\left(t_\II+ i \beta \delta\right)\right].
\end{array}
\right.
\end{eqnarray}
In terms of the real \EX variables, we have
\begin{eqnarray}\mlab{mex3}
\left\{
\begin{array}{rcl}
\etad &=& +\eta \, \cos\left(2\pi\delta\right)
+ i\xi\,\sin\left(2\pi\delta\right),  \\[2mm]
i\xid &=& -\eta \, \sin\left(2\pi\delta\right)
+ i\xi\,\cos\left(2\pi\delta\right),
\end{array}
\right.
\end{eqnarray}
or equivalently
\begin{eqnarray}\mlab{mex3b}
\left\{
\begin{array}{rcl}
\xid^+ &=& \exp\left(+i2\pi\delta\right)\,\xi^+, \\[2mm]
\xid^- &=& \exp\left(-i2\pi\delta\right)\,\xi^-.
\end{array}
\right. \label{defxipm}
\end{eqnarray}
The time shift induces thus a rotation in the $(\eta,i\xi)$ plane of
$R_\II$. The metric becomes in terms of the rotated coordinates
\begin{equation}\mlab{mex1}
ds^2 = \frac{-d\eta_\delta^2 +d\xi_\delta^2}
{\alpha^2\left(\xi_\delta^2-\eta^2_\delta\right)} + dy^2 + dz^2,
\end{equation}
and is thus unchanged by the time shift which is therefore an isometry
of the four-dimensional complex \EX spacetime.  The equations
(\ref{ACtx}) become after the time shift
\begin{eqnarray}
\left\{ \begin{array}{rcl}
t_\IId(\xid) &=& t_\I(\xid) +i\,\beta\left(1/2 + \delta\right),
\\[2mm]
x_\IId(\xid) &=& x_\I(\xid).
\end{array} \right.
\label{ACtxd}
\end{eqnarray}

\section{Fields in \EX spacetime}

We now consider a free scalar field in the \EX spacetime (for the case
of a fermion field see Ref.~\cite{GUI92ferm}).  The ``global'' scalar
field in \EX coordinates shall be denoted by $\Phi(\xi)$. It satisfies
to the Klein-Gordon equation
\begin{eqnarray} \mlab{kge}
(\Box + m^2) \Phi(\xi) = 0,
\end{eqnarray}
where $\Box = \nabla_\mu\nabla^\mu$ if $\nabla_\mu$ is the covariant
derivative.  The scalar product of two fields is given by
\begin{eqnarray}
(\Phi_1,\Phi_2)=
-i\int_\Sigma d\Sigma \parallel\!g(\xi)\!\parallel^{1/2}
\Phi_1(\xi)n^\nu\stackrel{\leftrightarrow}{\partial}_\nu \Phi_2^*(\xi),
\label{scalarproduct}
\end{eqnarray}
where $g$ is the determinant of the metric $g^{\mu\nu}$, 
$\Sigma$ is any space-like surface and $n^\nu$ an orthonormal
vector to this surface.

\subsection{Euclidean section}

In the Euclidean section we have $\Phi=\Phi(\sigma,\xi,y,z)$, and the
field in the $t$-$x$ coordinates defined in Eq.~(\ref{euclidTrans})
shall be denoted by $\phi=\phi(\tau,x,y,z)$. These two fields are
related by
\begin{equation}
\phi(\tau,x,y,z) =
\Phi(\sigma(\tau,x),\xi(\tau,x),y,z).
\end{equation}
Because of the periodic nature of the time $\tau$ and by
imposing single valuedness we have necessarily
\begin{equation}
\phi(\tau,x,y,z) = \phi(\tau+\beta,x,y,z).
\label{periodicfield}
\end{equation}

\subsection{Lorentzian section}

In the Lorentzian section, as we have seen, we have four different
regions, each of them being a complete Minkowski spacetime.  Since we
are interested only in regions $R_\I$ and $R_\II$, we shall consider
the quantum field over these two regions only.  Our aim is to find an
expansion for the global field $\Phi$ in the joining $R_\I \cup
R_\II$.

\smallskip
We start by defining the ``local'' fields $\phi^\I(x_\I)$ and
$\phi^\II(x_\II)$ by
\begin{eqnarray} \label{defphi}
\Phi(\xi) = \left\{
\begin{array}{lll}
\phi^\I(x_\I(\xi)), &\quad& \mbox{when $\xi\in R_\I$}, \\[2mm]
\phi^\II(x_\II(\xi)), && \mbox{when $\xi\in R_\II$}.
\end{array}\right.
\end{eqnarray}
They have support in $R_\I$ and $R_\II$ respectively. By choosing the
particular surface $\eta=a\xi$ where $a$ is a constant satisfying
$-1<a<1$, one shows from Eq.~(\ref{scalarproduct}) that the global
scalar product is given by
\begin{eqnarray}
(\Phi_1,\Phi_2)=\ <\phi^\I_1,\phi^\I_2> + <\phi^\II_1,\phi^\II_2>,
\label{SP}
\end{eqnarray}
where $<\,,\,>$ is the local scalar product in Minkowski spacetime
\begin{eqnarray}
<\phi_1,\phi_2>\ = -i\int_{\R^3} d^3x\,
\phi_1(x)\stackrel{\leftrightarrow}{\partial}_t\phi_2^*(x).
\label{SPM}
\end{eqnarray}

\bigskip
In \EX spacetime covered by $t$-$x$ coordinates given in
Eqs.~(\ref{TransI}) and (\ref{TransII}), the mode solutions of the
Klein-Gordon equation are just local plane waves restricted to a given
region. They are given by
\begin{eqnarray}
u_\Vk(x_\I) &=& \left(4\pi\omega_\Vk\right)^{-\frac{1}{2}}\,
e^{i\left(-\omega_\Vk\,t_\I+\Vk\cdot\Vx_\I\right)},
\mlab{MMu} \\[2mm]
v_\Vk(x_\II) &=& \left(4\pi\omega_\Vk\right)^{-\frac{1}{2}}\,
e^{i\left(+\omega_\Vk\,t_\II+\Vk\cdot\Vx_\II\right)},
\mlab{MMv}
\end{eqnarray}
where $\omega_\Vk= \sqrt{\Vk^2+m^2}$. From these Minkowski modes, one
defines the two wave functions $U_\Vk(\xi)$ and $V_\Vk(\xi)$ with
support in $R_\I$ and $R_\II$ respectively by
\begin{eqnarray}
U_\Vk(\xi) &=&  \left\{
\begin{array}{lcl}
u_\Vk(x_\I(\xi)), &\quad&
\mbox{when $\xi\in R_\I$}, \\[2mm]
0, && \mbox{when $\xi\in R_\II$},
\end{array}\right.
\label{defU}\\[2mm]
V_\Vk(\xi) &=&
\left\{\begin{array}{lcl}
0, && \mbox{when $\xi\in R_\I$}, \\[2mm]
v_\Vk(x_\II(\xi)), &\quad& \mbox{when $\xi\in R_\II$}.
\end{array}\right.
\label{defV}
\end{eqnarray}
Their power spectrum with respect to the momenta conjugated to $\xi^+$
and $\xi^-$ contains negative contributions, which are furthermore not
bounded by below. Consequently, the sets of functions $\left\{
U_\Vk(\xi), U^*_{-\Vk}(\xi)\right\}_{\Vk\in\R^3}$ and $\left\{
V_\Vk(\xi), V^*_{-\Vk}(\xi)\right\}_{\Vk\in\R^3}$ defined on $R_\I$
and $R_\II$ respectively are both over-complete since the same energy
contribution (i.e.~momentum contribution conjugate to $\eta$) can
appear twice in these sets. In other words, the energy spectrum of
$U_\Vk$ and $U^*_{-\Vk}$ overlap, and so do the ones of $V_\Vk$ and
$V^*_{-\Vk}$. These sets can thus not be used as a basis in their
respective regions, and the joining of these sets is clearly not a
basis in $R_\I\cup R_\II$.

\smallskip
To construct a basis in $R_\I\cup R_\II$, we could solve the
Klein-Gordon equations in \EX coordinates to obtain the field modes in
these coordinates. However, the Bogoliubov transformations resulting
from this basis choice are rather complicated. So instead of doing
this, we shall construct from the wave functions $u_\Vk(x_\I(\xi))$
and $v^*_{-\Vk}(x_\II(\xi))$ basis elements having positive energy
spectrum.

\smallskip
We shall demand these basis elements to be analytical functions in
the lower complex planes of $\xi^+$ and $\xi^-$, so that their
spectrum contain only positive contributions of the momenta conjugate
with respect to $\xi^+$ and $\xi^-$. In consequence, they shall have
positive energy spectra.

\smallskip
We thus extend analytically the two wave functions $u_\Vk(x_\I(\xi))$
and $v^*_{-\Vk}(x_\II(\xi))$ in the lower complex planes of $\xi^+$
and $\xi^-$ (as in Section \ref{subsecAC}, the cut in the complex
planes is given by $\R_-+i0^+$). By applying formula (\ref{ACtx})
we get directly
\begin{eqnarray}
u_\Vk(x_\I(\xi)) &=& e^{-\frac{\beta}{2}\omega_\Vk}\,v^*_{-\Vk}(x_\II(\xi)),
\label{ACuv} \\[2mm]
v_\Vk(x_\II(\xi)) &=& e^{-\frac{\beta}{2}\omega_\Vk}\,u^*_{-\Vk}(x_\I(\xi)).
\label{ACvu}
\end{eqnarray}
The expressions on the right and left hand sides of these last
equations are analytic continuations of each others. In this way we
are led to introduce the two normalized linear combinations
\begin{eqnarray}
\left\{ \begin{array}{rcl}
F_{\Vk}(\xi) &=& \left(1-f_\Vk\right)^{-\frac{1}{2}}
\left[ U_\Vk(\xi) + f_\Vk^{\frac{1}{2}}\,V^*_{-\Vk}(\xi)\right],
\\[3mm]
{\widetilde F}_{\Vk}(\xi) &=& \left(1-f_\Vk\right)^{-\frac{1}{2}}
\left[ V_{\Vk}(\xi) + f_\Vk^{\frac{1}{2}}\,U^*_{-\Vk}(\xi)\right],
\end{array} \right.
\label{globalmodes}
\end{eqnarray}
where $f_\Vk=e^{-\beta\omega_\Vk}$, and where $U_\Vk(\xi)$ and
$V_\Vk(\xi)$ are defined in Eqs.~(\ref{defU}) and (\ref{defV}). These
wave functions are still solutions of the Klein-Gordon equation. They
are analytical in $R_\I\cup R_\II$ and in particular at the origin
$\xi^+=\xi^-=0$.  Since they are analytical complex functions in the
lower complex planes of $\xi^+$ and $\xi^-$, their spectrum has only
positive energy contributions. The set $\{F_{\Vk}, F^*_{-\Vk},
{\widetilde F}_\Vk, {\widetilde F}^*_{-\Vk} \}_{\Vk\in\R^3}$ is thus
complete but not over-complete over the joining $R_\I\cup
R_\II$. Furthermore it is an orthogonal set since
\begin{eqnarray}
\begin{array}{rcccl}
(F_\Vk,F_\Vp)&=& (\widetilde{F}^*_\Vk,\widetilde{F}^*_\Vp)
&=& +\delta^3(\Vk-\Vp),
\\[2mm]
(F^*_\Vk,F^*_\Vp)&=& (\widetilde{F}_\Vk,\widetilde{F}_\Vp)
&=& -\delta^3(\Vk-\Vp),
\end{array}
\end{eqnarray}
with all the other scalar products vanishing.

\bigskip
On one hand, the local scalar fields can be expanded in the Minkowski
modes given in Eqs.~(\ref{MMu}) and (\ref{MMv}),
\begin{eqnarray}
\phi^\I(x_\I) &=& \int d^3\Vk \left[ a^\I_\Vk\,u_\Vk(x_\I)+
a^{\I\dagger}_\Vk\,u^*_\Vk(x_\I)\right],
\label{ExpI}\\[1mm]
\phi^\II(x_\II) &=& \int d^3\Vk \left[ a^\II_\Vk\,v_\Vk(x_\II)+
a^{_\II\dagger}_\Vk\,v^*_\Vk(x_\II)\right].
\label{ExpII}
\end{eqnarray}
And on the other hand, the global scalar field can be expanded in
terms of the ``global'' modes given in Eqs.~(\ref{globalmodes}) as
\begin{eqnarray}
\lefteqn{ \Phi(\xi)= \int d^3\Vk \left[
b_\Vk\,F_\Vk(\xi) + b^\dagger_\Vk\,F^*_{-\Vk}(\xi)
\right. } && \nonumber\\[0mm]&& \hspace{32mm}\left.
+\,\tilde{b}_\Vk\,\widetilde{F}_\Vk(\xi)
+\tilde{b}_\Vk^\dagger\,\widetilde{F}^*_{-\Vk}(\xi) \right].
\mlab{Exp}
\end{eqnarray}
These three expansions define the local and global creation and
annihilation operators, which are related by Bogoliubov
transformations. To obtain these, we recall the definition
(\ref{defphi}) relating the local and global fields and we use the
field expansions (\ref{ExpI}), (\ref{ExpII}) and (\ref{Exp}). We
obtain
\begin{eqnarray}\label{bogol}
\left\{ \begin{array}{rcl}
b_\Vk&=& a^\I_\Vk\,\cosh\theta_\Vk - a^{\II\dagger}_\Vk\,\sinh\theta_\Vk,
\\[2mm]
{\tilde b}_\Vk &=&
a^\II_\Vk\,\cosh\theta_\Vk-a^{\I\dagger}_\Vk\,\sinh\theta_\Vk,
\end{array} \right.
\end{eqnarray}
where $\sinh^2\theta_\Vk=n(\omega_\Vk)=(e^{\beta\omega_\Vk } -1)^{-1}$.

\subsection{Extended Lorentzian section}

By following the above procedure, we shall now construct a set of
positive energy modes defined in the extended Lorentzian section
introduced in Section \ref{sub:ELS}. We start by considering region
$R_\II$, where the set
$\{v_\Vk(x_\II),v^*_{-\Vk}(x_\II)\}_{\Vk\in\R^3}$ is a plane wave
basis. One has
\begin{eqnarray}
v_\Vk(x_\II) &=& \left(4\pi\omega_\Vk\right)^{-\frac{1}{2}}\,
e^{i\left(+\omega_\Vk\,t_\II+\Vk\cdot\Vx_\II\right)},
\\[2mm]
v_{-\Vk}^*(x_\II) &=& \left(4\pi\omega_\Vk\right)^{-\frac{1}{2}}\,
e^{i\left(-\omega_\Vk\,t_\II+\Vk\cdot\Vx_\II\right)}.
\end{eqnarray}
Under the time shift, Eq.~(\ref{shift}), this set is transformed into
$\{v_\Vk(x_{\IId}),v^\sharp_{-\Vk}(x_{\IId})\}_{\Vk\in\R^3}$, where we
have replaced the symbol $\ast$ by $\sharp$ because
$v^\sharp_{-\Vk}(x_{\IId})$ is no longer the complex conjugate of
$v_\Vk(x_{\IId})$. Indeed, one has
\begin{eqnarray}
v_\Vk(x_{\IId}(x_\II)) &=& e^{-\beta\omega_\Vk\delta}\,v_\Vk(x_\II),
\label{vII}\\[2mm]
v^\sharp_{-\Vk}(x_{\IId}(x_\II)) &=&
e^{+\beta\omega_\Vk\delta}\,v_{-\Vk}^*(x_\II).
\label{vIIs}
\end{eqnarray}
The complex conjugation and the time shift are not commutative.  We
notice that $v^\sharp_{-\Vk}(x_{\IId}(x_\II))$ can be obtained from
$v_\Vk(x_{\IId}(x_\II))$ by complex conjugation {\it and} by the
replacement $\delta\rightarrow-\delta$. This rule actually defines the
$\sharp$-conjugation. 

\smallskip
In the same way as in Eq.~(\ref{defphi}) one defines 
\begin{eqnarray} \label{defphid}
\Phi(\xi) = \left\{
\begin{array}{lll}
\phi^\I(x_\I(\xi)), &\quad& \mbox{when $\xi\in R_\I$}, \\[2mm]
\phi^\IId(x_\IId(\xi)), && \mbox{when $\xi\in R_\IId$}.
\end{array}\right.
\end{eqnarray}
The global scalar product, Eq.~(\ref{SP}), is modified and becomes in
$R_\I\cup R_\IId$,
\begin{eqnarray}
(\Phi_1,\Phi_2)_{_\delta} = \
<\phi^\I_1,\phi^\I_2> + <\phi^\IId_1,\phi^\IId_2>_{_\delta},
\end{eqnarray}
where the local Minkowski scalar product $<\ ,\ >_{_\delta}$ in region
$R_\IId$ is given by
\begin{eqnarray}
<\phi_1,\phi_2>_{_\delta}\ =-i\int_{\R^3} d^3x_\IId\,
\phi_1(x_\IId)\stackrel{\leftrightarrow}{\partial}_t \phi_2^\sharp(x_\IId).
\end{eqnarray}

\bigskip
Equations (\ref{ACuv}), (\ref{ACvu}), (\ref{vII}) and (\ref{vIIs})
imply
\begin{eqnarray}
u_\Vk(x_\I(\xi_\delta)) &=&
e^{-\beta\omega_\Vk(1/2+\delta)}\,v^\sharp_{-\Vk}(x_{\IId}(\xi_\delta)),
\label{ACuvd} \\[2mm]
v_\Vk(x_\IId(\xi_\delta)) &=&
e^{-\beta\omega_\Vk(1/2+\delta)}\,u^\sharp_{-\Vk}(x_\I(\xi_\delta)).
\label{ACvud}
\end{eqnarray}
Consequently one also has
\begin{eqnarray}
u^\sharp_\Vk(x_\I(\xi_\delta)) &=& e^{-\beta\omega_\Vk(1/2-\delta)}\,
v_{-\Vk}(x_{\IId}(\xi_\delta)),
\label{ACuvds}\\[2mm]
v^\sharp_\Vk(x_\IId(\xi_\delta)) &=&
e^{-\beta\omega_\Vk(1/2-\delta)}\,u_{-\Vk}(x_\I(\xi_\delta)).
\label{ACvuds}
\end{eqnarray}
The expressions on the left and right hand sides of these equations
are thus analytic continuations of each others. If we now define
\begin{eqnarray}
U_\Vk(\xid) &=&
\left\{\begin{array}{lcl} 
u_\Vk(x_\I(\xid)), &\quad&
\mbox{when $\xid\in R_I$}, \\[2mm]
0, && \mbox{when $\xid\in R_\IId$},
\end{array}\right.
\\[3mm]
V_\Vk(\xid) &=&
\left\{\begin{array}{lcl}
0, && \mbox{when $\xid\in R_I$}, \\[2mm]
v_\Vk(x_{\IId}(\xid)), &\quad&
\mbox{when $\xid\in R_\IId$},
\end{array}\right.
\end{eqnarray}
the global modes in the extended Lorentzian section $R_\I\cup R_\IId$ are
given by
\begin{eqnarray} \mlab{ex11}
\left\{ \begin{array}{rcl}
G_{\Vk}(\xi_\delta) &=& (1-f_\Vk)^{-\frac{1}{2}} \left[ U_\Vk(\xi_\delta)
+f_\Vk^{\frac{1}{2}+\delta}\,V^\sharp_{-\Vk}(\xi_\delta) \right],
\\[2mm]
{\widetilde G}_{\Vk}(\xi_\delta) &=&
(1-f_\Vk)^{-\frac{1}{2}}\left[ V_\Vk(\xi_\delta) +
f_\Vk^{\frac{1}{2}+\delta}\,U^\sharp_{-\Vk}(\xi_\delta) \right],
\end{array}\right.
\label{GM}
\end{eqnarray}
and
\begin{eqnarray} \mlab{ex11bis}
\left\{ \begin{array}{rcl}
G^\sharp_{\Vk}(\xi_\delta) &=&
(1-f_\Vk)^{-\frac{1}{2}} \left[ U^\sharp_\Vk(\xi_\delta) +
f_\Vk^{\frac{1}{2}-\delta}\,V_{-\Vk}(\xi_\delta) \right],
\\[2mm]
{\widetilde G}^\sharp_{\Vk}(\xi_\delta) &=&
(1-f_\Vk)^{-\frac{1}{2}} \left[ V^\sharp_\Vk(\xi_\delta)+
f_\Vk^{\frac{1}{2}-\delta}\,U_{-\Vk}(\xi_\delta) \right],
\end{array} \right.
\label{GMs}
\end{eqnarray}
where $f_\Vk=e^{-\beta\omega_\Vk}$. From Eqs.~(\ref{ACuvd}) to
(\ref{ACvuds}), we see that the global modes are analytic in $R_\I\cup
R_\IId$, in particular at the origin $\xi^+_\delta=\xi^-_\delta=0$.
Since they are analytical complex functions on the lower complex
planes of $\xi^+_\delta$ and $\xi^-_\delta$, their spectrum has only
positive energy contributions.  When $\delta=0$, the above
combinations consistently reduce to the expressions of
Eq.~(\ref{globalmodes}).

\smallskip
We emphasize that the global modes $G^*_\Vk$ and $\widetilde{G}^*_\Vk$
are {\em not} analytic in the extended Lorentzian section, contrary to
the the non-hermitian combinations $G^\sharp_\Vk$ and
$\widetilde{G}^\sharp_\Vk$.  Non-hermitian conjugation operations such
as our sharp conjugation $\sharp$ are actually common in TFT, see for
example Ref.~\cite{BESV98} for a formally similar situation. In
Ref.~\cite{OST} it is shown the necessity of the so-called
Osterwalder-Schrader reflection as opposed to the hermiticity property in
Euclidean field theories even when the temperature vanishes.

\smallskip
The set $\{G_{\Vk},{\widetilde G}_{\Vk},G^\sharp_{\Vk}, {\widetilde
G}^\sharp_{\Vk}\}_{\Vk\in\R^3}$ is thus complete over $R_\I\cup
R_\IId$. It is furthermore an orthogonal set since
\begin{eqnarray}
\begin{array}{rcccl}
(G_\Vk,G_\Vp)_\delta &=&
(\widetilde{G}^\sharp_\Vk,\widetilde{G}^\sharp_\Vp)_\delta
&=& +\delta^3(\Vk-\Vp), \\[2mm]
(G^\sharp_\Vk,G^\sharp_\Vp)_\delta &=&
(\widetilde{G}_\Vk,\widetilde{G}_\Vp)_\delta
&=& -\delta^3(\Vk-\Vp),
\end{array}
\end{eqnarray}
where all the other scalar products vanish.

\bigskip
On one hand, the expansions of the local fields in the Minkowski modes
over regions $R_\I$ and $R_\IId$ are given respectively by
\begin{eqnarray} \mlab{expansionydelta1}
\phi^\I(x_\I) &=&
\int d^3\Vk \left[ a^\I_\Vk\,u_\Vk(x_\I)+
a^{\I\dagger}_\Vk\,u^*_\Vk(x_\I)\right],
\\[1mm]\mlab{expansionydelta2}
\phi^\IId(x_\IId) &=&
\int d^3\Vk \left[ a^\IId_\Vk\,v_\Vk(x_\IId)+
a^{_\IId\dagger}_\Vk\,v^*_\Vk(x_\IId)\right].
\end{eqnarray}
On the other hand, the expansion of the global field in the $G$ modes
over the joining $R_\I\cup R_\IId$ is
\begin{eqnarray}
\lefteqn{\Phi(\xi_\delta) = \int d^3\Vk\left[ c_\Vk\, G_\Vk(\xi_\delta)
+ c^\sharp_\Vk\,G^\sharp_{\Vk}(\xi_\delta) \right. }
&& \nonumber \\ && \hspace{32mm}\left.
+\ {\tilde c}_\Vk\,{\tilde G}_{-\Vk}(\xi_\delta)\,
+ {\tilde c}^\sharp_\Vk\,{\tilde G}^\sharp_{-\Vk}(\xi_\delta) \right].
\mlab{ex12}
\end{eqnarray}
{}From these last expansions and by using Eqs.~(\ref{GM}) and
(\ref{GMs}), one finds the Bogoliubov transformations,
\begin{eqnarray}
\left\{\begin{array}{lcr}
c_\Vk &=& (1-f_\Vk)^{-\frac{1}{2}}
\left(a^\I_\Vk-f_\Vk^{\frac{1}{2}-\delta}\,a^{\IId\dag}_\Vk\right),
\\[3mm]
{\tilde c}_\Vk &=& (1-f_\Vk)^{-\frac{1}{2}}
\left(a^\IId_\Vk-f_\Vk^{\frac{1}{2}-\delta}\,a^{\I\dag}_\Vk\right),
\end{array}\right.
\label{BEc}
\end{eqnarray}
and their $\sharp$-conjugate equivalent,
\begin{eqnarray}
\left\{\begin{array}{lcr}
c^\sharp_\Vk &=& (1-f_\Vk)^{-\frac{1}{2}}
\left( a^{\I\dag}_\Vk - f_\Vk^{\frac{1}{2}+\delta}\,a^\IId_\Vk\right),
\\[3mm]
{\tilde c}^\sharp_\Vk &=& (1-f_\Vk)^{-\frac{1}{2}}
\left( a^{\IId\dag}_\Vk-f_\Vk^{\frac{1}{2}+\delta}\,a^\I_\Vk\right).
\end{array}\right.
\label{BEcs}
\end{eqnarray}
Again the transformations given in Eq.~(\ref{bogol}) are recovered
when $\delta=0$.

\section{Thermal field theories in \EX spacetime}

We now show the equivalence of Quantum Field Theory in \EX spacetime
with Thermal Field Theories by considering the scalar field. We shall
see that in several sections of \EX spacetime QFT naturally reproduces
the known formalisms of TFTs insofar as the correct thermal Green
functions are recovered there.

\smallskip
On one hand, we shall show that in the Euclidean section of \EX
spacetime QFT corresponds to the imaginary time formalism, and that
the Green functions are the Matsubara Green functions in this section.
On the other hand, we shall show that in the extended Lorentzian
section QFT reproduces the two known formalisms of TFTs with real
time, namely the POM formalism and TFD.  Furthermore we shall fix the
parameter $\delta$ of the extended Lorentzian section with respect to
the parameter $\sigma$ of the POM formalism and of TFD. The most
general thermal matrix propagator shall be obtained in the framework
of \EX spacetime. We shall need below the covariant Lagrangian for the
scalar field with a source term ${\rm J}$. It is given by
\begin{equation}
{\cal L}[\Phi,{\rm J}] = \sqrt{-g} \left(
\frac{1}{2}\,\partial_\mu\Phi\partial_\nu\Phi +
\frac{m^2}{2}\,\Phi^2 - V(\Phi) - {\rm J} \Phi\right).
\end{equation}

\subsection{Imaginary time formalism}

The generating functional for the Green functions in the Euclidean
section of \EX spacetime is \cite{GUI90}
\begin{equation}
Z_E[J] = N \int [d\Phi] \exp\left\{-\int d\sigma d\xi dydz\
{\cal L}_{\sigma,\xi}[\Phi,{\rm J}]\right\},
\end{equation}
where
\begin{eqnarray}
\lefteqn{
{\cal L}_{\sigma,\xi}[\Phi,{\rm J}] = 
\frac{1}{2}\,\left[\left(\partial_\sigma \Phi\right)^2 
+ \left(\partial_\xi \Phi\right)^2\right] 
} \mlab{euclexz} \\ \nonumber &&\qquad 
+\, \frac{1}{\alpha^2\left(\sigma^2 +\xi^2\right)} \left\{
\frac{1}{2}\left(\nabla_{\!\perp}\Phi\right)^2
+ \frac{m^2}{2}\,\Phi^2 + V(\Phi) - {\rm J}\Phi \right\}.
\end{eqnarray}
We now perform the change of coordinates given in
Eq.~(\ref{euclidTrans}) in the generating functional. In the
$\tau$-$x$ coordinates, the sum over fields is taken over all periodic
fields satisfying to the periodic constraint in
Eq.~(\ref{periodicfield}),
\begin{equation}\mlab{euclz}
Z_E[J] = N \!\!\!\!\!\!\underset{\beta-periodic}{\int}\!\!\!\!\!\![d\phi] 
\exp\left\{ \!-\!\int_0^\beta\!\!\!\!d\tau\int_{\R^3}\!\!\!\!
dxdydz\,{\cal  L}_{\tau,x}[\phi,J] \right\}
\end{equation}
where $J(\tau,x,y,z)={\rm J}(\sigma,\xi,y,z)$ and
\begin{equation}\mlab{euclzL}
{\cal L}_{\tau,x}[\phi,J] 
= \frac{1}{2} \left[ \left( \partial_\tau\phi \right)^2 
+ \left( \nabla\phi \right)^2 
+ m^2\,\phi^2 \right] + V(\phi) - J\phi.
\end{equation}
By differentiation this last equation with respect to the source $J$
we obtain the Matsubara propagator whose Fourier transforms is
\begin{equation}
G_\beta({\bf k}, \omega_n) = \frac{1}{\omega_n^2 + {\bf k}^2 + m^2},
\end{equation}
where the Matsubara frequencies $\omega_n$ are given for bosons by
$\omega_n = 2\pi n/\beta$ ($n\in{\Bbb N}$).

\subsection{Real time formalism: Path Ordered Method}

Let us now consider the extended Lorentzian section of \EX spacetime.
The generating functional is given by \cite{GUI90}
\begin{equation}\mlab{lorentzexz}
Z[J] = {\cal N} \!\!\int [d\Phi] \exp \left\{\!i\int 
d\etad d\xid dydz\, {\cal L}_{\etad,\xid}[\Phi,{\rm J}] \right\},
\end{equation}
where
\begin{eqnarray}
\lefteqn{
{\cal L}_{\etad,\xid}[\Phi,{\rm J}] =
\frac{1}{2} \left[ \left( \partial_{\etad}\!\Phi \right)^2 
- \left( \partial_{\xid}\!\Phi \right)^2 \right] 
} && \label{lorentzact} \\[2mm] \nonumber 
&& + \, \frac{1}{\alpha^2\left(\etad^2-\xid^2\right)}
\left\{ - \frac{1}{2}\left(\nabla_{\!\perp}\Phi\right)^2 
-\frac{m^2}{2}\Phi^2 - V(\Phi) + {\rm J}\Phi \right\}.
\end{eqnarray}
Since we shall be interested only in the propagators whose spacetime
arguments belong only to the joining $R_\I\cup R_\IId$, we can set the
source to zero in regions $R_\III$ and $R_\IV$,
\begin{equation}\mlab{source}
{\rm J}(x) = 0, \qquad\qquad \mbox{when $x\in R_\III\cup R_\IV$}.
\end{equation}
Then the expression in Eq.~(\ref{lorentzexz}) reduces to
\begin{equation}\mlab{lorentzexz2}
Z[J] = {\cal N} \int [d\Phi] \exp \left\{ i\int_{R_\I\cup R_\IId}
\!\!\!\!\!\!\!\!\!\! d\etad d\xid dydz\
{\cal L}_{\etad,\xid}[\Phi,{\rm J}] \right\}.
\end{equation}
We now express the fields in regions $R_\I$ and $R_\IId$ in terms of
the local Minkowskian coordinates by using the transformations given
in Eq.~(\ref{mex2a}). We obtain
\begin{eqnarray}\mlab{lorentzexz3}
Z[J] &=& {\cal N} \int [d\phi] \exp \left\{i\int dt_\I dx_\I dydz\
{\cal L}_{t,x}[\phi,J] 
\right.  \\ \nonumber && \left. \qquad\qquad\qquad
+ \, i\int dt_{\IId} dx_{\IId} dydz\ {\cal L}_{t,x}[\phi,J] \right\},
\end{eqnarray}
where the integration is taken over the Minkowski spacetime, $\phi$ is
the local field and where
\begin{equation}\mlab{lorentzL}
{\cal L}_{t,x}[\phi,J] 
= \frac{1}{2} \left[ \left( \partial_t\phi \right)^2 
- \left( \nabla\phi \right)^2
- m^2\,\phi^2 \right] - V(\phi) + J\phi.
\end{equation}
We now use the relations (\ref{ACtxd}) and obtain
\begin{eqnarray}
Z[J] &=& {\cal N} \int [d\phi] \exp 
\bigg\{ i\int dtdxdydz 
\mlab{lorentzexz4}\\ \nonumber && \times
\big[ \, {\cal L}_{t,x}[\phi,J]\,(t,x)
- {\cal L}_{t,x}[\phi,J]\,(t + i\beta \delta,x) \, \big] \bigg\},
\end{eqnarray}
where in the last step we have dropped the subscript $I$ and where we
have taken into account the fact that the direction of time in
$R_\IId$ is opposite to the one in $R_\I$, resulting in a minus sign
in the second integration.

\smallskip
We now consider the expression for the generating functional as given
in the POM formalism \cite{LEB}. It is given by
\begin{equation}\mlab{POMgf}
Z_{\mbox{\tiny POM}}[J] = {\cal N}' \int [d\phi] 
\exp \left\{ i\int_C dtdxdydz\ {\cal L}_{t,x}[\phi,J]\right\},
\end{equation}
where the time path $C$ is shown in Fig.~\ref{fig1}.
The contribution from the
vertical parts of the contour may be included in the normalization
factor ${\cal N}'$ and neglected when calculating the real-time Green
functions.  The generating functionals in Eq.~(\ref{lorentzexz4}) and
(\ref{POMgf}) can then be identified provided that
\begin{eqnarray}\mlab{sigma}
\delta & = &  \sigma - 1/2.
\end{eqnarray}
We thus see that the type of time path in the POM formalism is related
directly to the ``rotation angle'' between the two regions $R_\I$ and
$R_\IId$ of \EX spacetime.
In the free field case, from the above generating functionals 
we obtain the well-known thermal
matrix propagator \cite{DAS,LEB,LW87},
\begin{eqnarray}
\begin{array}{rcl}
D_{11}(k) &=& \displaystyle
\frac{i}{k^2-m^2+i0^+}\, + 2\pi\,n(k_0)\,\delta(k^2-m^2),
\\[4mm]
D_{22}(k) &=& D_{11}^*(k),
\\[3mm]
D_{12}(k) &=& e^{\sigma\beta k_0}
\left[\,n(k_0)+\theta(-k_0)\,\right] 2\pi\,\delta(k^2-m^2),
\\[3mm]
D_{21}(k) &=& e^{-\sigma\beta k_0}
\left[\,n(k_0)+\theta(k_0)\,\right] 2\pi\,\delta(k^2-m^2),
\end{array}
\mlab{matprop}
\end{eqnarray}
where $n(k_0)=(e^{\beta |k_0|} -1)^{-1}$. The parameter $\sigma$
appears explicitly only in the off-diagonal components of the matrix
propagator.

\subsection{Real time formalism: Thermo Field Dynamics}

Another formalism for real-time TFT is Thermo Field Dynamics
\cite{UM0,UM1}.  In this approach a central role is played by the
Bogoliubov transformation, relating the zero-temperature annihilation
and creation operators with the thermal ones.  In TFD the field
algebra is doubled and one then considers two commuting field operators
$\phi$ and ${\tilde \phi}$ given by
\begin{equation}\mlab{4.21a}
\phi(x)=\int \, 
\frac{d^{3}{\bf k}}{(2\pi)^{\frac{3}{2}}
(2\omega_k)^{\frac{1}{2}}}\,
\left[a_{{\bf k}}e^{i\left(-\omega_k t + {\bf kx}\right)} \, + \, 
a^{\dag}_{{\bf k}}e^{i\left(\omega_k t-{\bf kx}\right)}\right],
\end{equation}
\begin{eqnarray}
\nonumber \\[-15mm]\nonumber
\end{eqnarray}
\begin{equation} \mlab{4.21b} 
{\tilde \phi}(x)=\int \, 
\frac{d^{3}{\bf k}}{(2\pi)^{\frac{3}{2}}
(2\omega_k)^{\frac{1}{2}}}\,
\left[{\tilde a}_{{\bf k}}e^{i\left(\omega_k t-{\bf kx}\right)}\,+
\, {\tilde a}^{\dag}_{{\bf k}} 
e^{i\left(-\omega_k t +{\bf kx}\right)}\right].
\end{equation}
In the original formulation of Takahashi and Umezawa \cite{UM0}, the
thermal Bogoliubov transformation is given by
\begin{eqnarray}
\left\{\begin{array}{lrc}
a_\Vk(\theta) &=& a_\Vk\cosh\theta_\Vk - {\tilde
a}^{\dag}_\Vk\sinh\theta_\Vk,
\\[2mm]
{\tilde a}_\Vk(\theta)
&=& {\tilde a}_\Vk\cosh\theta_\Vk - a^{\dag}_\Vk\sinh\theta_\Vk,
\end{array}\right.
\mlab{ex2}
\end{eqnarray}
where $\sinh^2\theta_{{\bf k}}=n(\omega_k)$.  These operators
annihilate the ``thermal vacuum'' $|0(\beta)\rangle$. Thermal averages
are calculated as expectation values with respect to
$|0(\beta)\rangle$.

\smallskip
The form of the thermal Bogoliubov matrix is however not unique. It is
possible to generalize the above transformation to a non-hermitian
superposition of the form \cite{UM1,H95}
\begin{eqnarray}
\left\{\begin{array}{rcl}
\gamma_\Vk &=& (1-f_\Vk)^{-\frac{1}{2}}
\left(a_\Vk-f_\Vk^{1-\sigma}\,{\tilde a}^\dagger_\Vk \right),
\\[2mm]
{\tilde\gamma}_\Vk &=& (1-f_\Vk)^{-\frac{1}{2}}
\left({\tilde a}_\Vk-f_\Vk^{1-\sigma}\,a^\dagger_\Vk \right),
\end{array} \right.
\mlab{ex3}
\end{eqnarray}
where $f_\Vk=e^{-\beta\omega_\Vk}$. The non-hermiticity property of
this last transformation implies that the canonical conjugates of
$\gamma$ and ${\tilde\gamma}$ are not $\gamma^\dagger$ and
${\tilde\gamma}^\dagger$, but are rather the combinations
\begin{eqnarray}
\left\{\begin{array}{rcl}
\gamma^\sharp_\Vk &=&(1-f_\Vk)^{-\frac{1}{2}}
\left( a^\dagger_\Vk -f_\Vk^\sigma\,{\tilde a}_\Vk \right),
\\[2mm]
{\tilde\gamma}^\sharp_\Vk &=& (1-f_\Vk)^{-\frac{1}{2}}
\left( {\tilde a}^\dagger_\Vk -f_\Vk^\sigma\, a_\Vk \right),
\end{array}\right.
\mlab{ex4}
\end{eqnarray}
which give the correct canonical commutators,
$[\gamma_\Vk,\gamma^\sharp_{\bf p}]=\delta^3({\bf k}-{\bf p})$, etc.  
Here
the $\sharp$ conjugation corresponds to the usual hermitian
conjugation $\dagger$ {\em together} with the replacement
$\sigma\rightarrow 1-\sigma$. The hermitian representation in
Eq.~(\ref{ex2}) is recovered when $\sigma = 1/2$.

\smallskip
Thermal averages are now expressed as \cite{UM1,H95}
\begin{equation}\mlab{4.31}
\langle A \rangle = 
\frac{\,_{_L}\!\langle 0(\beta)| A | 0(\beta)\rangle_{_R}}
{\,_{_L}\!\langle 0(\beta)|0(\beta)\rangle_{_R}},
\end{equation}
where $A$ is an observable, and where $| 0(\beta)\rangle_{_R}$ and
$\,_{_L}\!\langle 0(\beta)|$ are the left and right vacuum states
defined by
\begin{equation}\mlab{4.35}
\left.\begin{array}{c}
\gamma\\{\tilde \gamma}
\end{array}\right\}
|0(\beta)\rangle_{_R}=\,0 \,= 
\,_{_L}\!\langle 0(\beta)|
\left\{\begin{array}{c}
\gamma^{\sharp} \\{\tilde \gamma}^{\sharp}
\end{array}\right.
\end{equation}
The thermal propagator for a scalar field is calculated in TFD as
follows:
\begin{equation}\mlab{4.36}
D^{(ab)}(x,y) = 
\langle T\left[ \phi^a(x) \phi^b(y)^\dagger \right] \rangle,
\end{equation}
where 
\begin{equation}
\phi^a\equiv \left(\begin{array}{c}
\phi \\ {\tilde \phi} 
\end{array}\right).
\end{equation}
This propagator coincides with the one of Eqs.~(\ref{matprop}), as can
be easily checked by using the definitions given above.  The
connection of TFD with the geometrical picture of \EX spacetime is
immediate if we make the identification
\begin{equation}
\left(\begin{array}{c}
\phi \\ {\tilde \phi} 
\end{array}\right)
\equiv
\left(\begin{array}{c}
\phi^\I \\ \phi^\IId 
\end{array}\right)
\end{equation}
under the constraint of Eq.~(\ref{sigma}) (see also
Eq.~(\ref{defphid})). Then the Bogoliubov transformations in
Eqs.~(\ref{BEc}), (\ref{BEcs}) and those in Eqs.~(\ref{ex3}), (\ref{ex4})
are identical.

\section{Other features of the extended Lorentzian section}

In this Section we discuss other features of the \EX spacetime, which
are influenced by the rotation of Eq.~(\ref{mex3}).  Along the lines
of Ref.~\cite{ZG98}, we consider the analytic continuation of the
imaginary time thermal propagator to real times in the context of \EX
spacetime. In Ref.~\cite{ZG98} it was shown that the geometric
structure of \EX spacetime plays a crucial role in obtaining the
matrix real-time propagator from the Matsubara one.

\smallskip
To see how this works, it is sufficient to consider the simple case of
a massless free scalar field in two-dimensions. In the Euclidean
section of \EX spacetime, the equation for the propagator is
\begin{equation}\mlab{of1}
\left(\frac{\partial^2}{\partial \sigma^2} 
+\frac{\partial^2}{\partial \xi^2} \right) D_I(A-A') 
= - (-g_E)^{-\frac{1}{2}}\delta^2(A-A'),
\end{equation}
where $(A,A')$ denotes a couple of points, $g_E$ is the determinant of
the metric in the Euclidean section, and where $D_I$ is the
imaginary time thermal propagator. Now we continue Eq.~(\ref{of1}) to the
extended Lorentzian section. This is achieved by first replacing
$\sigma$ by $i\eta$ and then performing the rotation of
Eq.~(\ref{mex3}). If $D$ is the real time propagator, we have
\begin{equation}\mlab{of2}
\left(-\frac{\partial^2}{\partial \etad^{2}} +
\frac{\partial^2}{\partial \xid^{2}} \right) D(A-A') 
= - (-g_{L})^{-\frac{1}{2}}\delta^2(A-A'),
\end{equation}
where $g_{L}$ stands for the determinant of the metric in the
Lorentzian section. Because of the presence of different disconnected
regions in the Lorentzian section, the propagator has a matrix
structure, since now the points $A$ and $A'$ can belong either to
region $R_{\I}$ or $R_{\IId}$ ($R_{\III}$ and $R_{\IV}$ are excluded
since they are space-like with respect to $R_{\I}$ and
$R_{\IId}$). The delta function with complex arguments $A$ and $A'$ is
defined to be zero when the points belong to different regions.

\smallskip
In Minkowski coordinates, Eq.~(\ref{of2}) reads
\begin{equation}\mlab{of3}
\left(-\frac{\partial^2}{\partial t^{2}} 
+ \frac{\partial^2}{\partial x^{2}} \right) D(A - A') 
= - \delta^2_C(A - A').
\end{equation}
The delta function containing the time variable in the last equation
needs to be defined on an appropriate path since the time variable is
complex. This path coincides with the POM time path of Fig.~\ref{fig1}
when the identification of Eq.~(\ref{sigma}) is made.  By use of
Eq.~(\ref{ACtxd}) and by following the procedure of Ref.~\cite{ZG98}
we obtain, for example for the component $D_{12}$, the equation
\begin{eqnarray}
\lefteqn{ \left(-\frac{\partial^2}{\partial t^{2}}
+\frac{\partial^2}{\partial x^{2}}\right)D(t-t'+i\sigma\beta,x -x') }
&& \nonumber \\ && \hspace{30mm}
= - \delta_C(t-t'+i\sigma\beta) \,\delta(x -x'),
\mlab{of5}
\end{eqnarray}
which has the $D_{12}$ propagator given in Eq.~(\ref{matprop}) as
solution. 

\bigskip
Let us finally consider the {\em tilde conjugation} in the context of
\EX spacetime. The tilde conjugation rules are postulated in TFD in
order to connect the physical and the tilde operators.  Due to the
geometrical structure of \EX spacetime, these rules are there seen as
coordinate transformations. This was first discussed in
Ref.~\cite{ZG95}. We extend here the result to the extended Lorentzian
section of \EX spacetime.

\smallskip
Let us recall the tilde rules as defined in TFD \cite{UM0,UM1} (we
restrict for simplicity to bosonic operators):
\begin{eqnarray}
\begin{array}{rclcrcl}
\left(A B\right)\tilde{} &=& \tilde{A} \tilde{B}, &&
\left(C_1 A + C_2 B \right)\tilde{} &=& C_1^*\tilde{A} + C_2^*\tilde{B},
\\
\left(\tilde{A}\right)\tilde{} &=& A, &&
\left(\tilde{A}\right)^\dagger &=& \left(A^\dagger\right)\tilde{},
\end{array}\mlab{of6}
\end{eqnarray}
where $A, B$ are operators and $C_1, C_2$ are c-numbers.  In order to
reproduce this operation in the extended Lorentzian section of \EX
spacetime, let us first introduce the following $M$ operation as
defined in Ref.\cite{ZG95}.
\begin{equation}\mlab{of7}
M\,\Phi(\eta, \xi)\,M^{-1} \equiv \Phi(-\eta, -\xi).
\end{equation}
When the field is expressed in Minkowskian coordinates the $M$
operation reads as
\begin{equation}\mlab{of8}
M \,\phi(t, x)\, M^{-1} = \phi(t-i\beta/2, x).
\end{equation}
The $M$ operation is anti-linear, since it induces a time inversion
together with the shift $t\rightarrow t-i\beta/2$, see
Ref.\cite{ZG95}. This is clear when we consider its action on the
conjugate momentum $\pi(t,x)\equiv \partial_t \phi^\dagger(t,x)$,
\begin{equation}\mlab{of8b}
M\,\pi(t, x)\,M^{-1} = - \pi(t-i\beta/2, x).
\end{equation}
Next we perform a rotation by an angle $\delta$ transforming the
$\eta,\xi$ coordinates according to Eqs.~(\ref{mex3}) and the field
becomes then
\begin{equation}\mlab{of8c}
R_\delta\,\Phi(\eta,\xi)\,R_\delta^{-1} \equiv \Phi(\etad,\xid).
\end{equation}

\smallskip
Finally we introduce a $\delta$ conjugation operation, which is
similar to a charge conjugation, by
\begin{equation}\mlab{of9}
C_\delta\,\phi(t, x)\,C_\delta^{-1} \equiv \phi^\sharp(t, x),
\end{equation}
where the change $\delta\rightarrow -\delta$ (or equivalently
$\sigma\rightarrow 1-\sigma $ ) has to be performed together with
usual charge conjugation.

\smallskip
The combined action of the three operations results in the tilde
conjugation. By defining for simplicity the notation $G_\delta \equiv
C_\delta\,R_\delta\,M$, we get
\begin{eqnarray}\mlab{of10}
\begin{array}{rcl}
G_\delta\,\phi(t, x)\,G_\delta^{-1} 
&=&  \phi^\sharp(t-i\sigma \beta, x)
\nonumber \\[1mm] 
&=& \phi^\dagger(t-i(1-\sigma)\beta,x).
\end{array}
\end{eqnarray}
We then get by omitting the space dependence for simplicity
\begin{eqnarray} \nonumber
G_\delta\, \phi_1(t)\,\phi_2(t')\,G_\delta^{-1}
&=& \phi_1^\sharp(t-i\sigma \beta)\,\phi_2^\sharp(t'-i\sigma\beta),
\\[2mm] \nonumber
G_\delta \left[G_\delta\,\phi(t)\,G_\delta^{-1}\right] G_\delta^{-1}
&=& \phi(t),
\\[2mm] \mlab{of11}
G_\delta \left[ B_1\,\phi_1(t) + B_2\,\phi_2(t') \right]
G_\delta^{-1} &&
\\[1mm] =
B_1^*\,\phi_1^\sharp(t&-&i\sigma \beta) +
B_2^*\,\phi_2^\sharp(t'-i\sigma\beta),
\nonumber \\[2mm] \nonumber
G_\delta\left[ \phi^\dagger(t) \right] G_\delta^{-1} 
&=& \left[ G_\delta\,\phi(t)\,G_\delta^{-1}\right]^\dagger.
\end{eqnarray}
The c-numbers are conjugated since the $M$ operation is
anti-linear. The second of the above relations is derived as follows
\begin{eqnarray}
\lefteqn{ G_\delta \left[G_\delta \,
\phi(t)\, G_\delta^{-1}\right] G_\delta^{-1} }
&& \nonumber \\[1mm] \nonumber
&& = C_\delta\,R_\delta\,M\,\phi^\dagger(t-i(1-\sigma) \beta)\,
M^{-1}\,R_\delta^{-1}\, C_\delta^{-1}
\\[1mm] \nonumber
&&= C_\delta\,\phi^\dagger(t-i(1-\sigma)\beta - i\sigma\beta)
\, C_\delta^{-1}
\\[2mm] \mlab{of12}
&& = \phi(t-i\beta) = \phi(t).
\end{eqnarray}
The tilde rules of Eq.~(\ref{of6}) are thus reproduced.

\section{\bf Discussion and Conclusions}

We have discussed a section of \EX spacetime which represent the
general geometric background for real-time thermal field theories at
equilibrium. This section is complex and can be regarded as an
extension of the usual Lorentzian section of \EX spacetime by means of
a rotation of region $R_\II$ with respect to $R_\I$ in the complex \EX
spacetime. In terms of Minkowski coordinates the rotation is
equivalent to a constant time shift, leaving the metric invariant.

\smallskip
The angle between the two regions turns out to be related to the
$\sigma$ parameter of the time path as used in the POM formalism. It
also reproduces the parameter present in the Bogoliubov thermal matrix
of TFD, when the relation between modes belonging to different regions
is considered.  The general form of the thermal matrix propagator
containing the parameter $\sigma$ has been obtained by use of this
particular geometrical background.  Finally, we have discussed the
analytic continuation of imaginary time propagator to real time matrix
propagator and the tilde rule in the context of \EX spacetime.

\smallskip
In the geometrical background of \EX spacetime, it is possible to
understand the differences between the various formalisms of TFT in a
simple way. In particular, with regards to the real-time methods,
i.e.~the POM formalism and TFD, the geometric picture is the
following. In the Lorentzian section of \EX spacetime we have two
different regions $R_\I$ and $R_\IId$ over which the field is defined.
For a global observer this field is a free field. However when one
does restrict to one of the two regions (say $R_\I$), temperature
arise as a consequence of the loss of information (increase in
entropy) about the other region.  In order to calculate the
propagator, one needs then to compare the fields defined in different
regions. This can be done essentially in two ways:

\smallskip
1) By  analytically continuing the field $\phi^\IId(x_\IId)$ defined in
region $R_\IId$ to region $R_\I$. Then the time argument gets shifted
by $i\beta(1/2+\delta)$, as described in Section III. One thus ends up
with {\em one field} and  {\em two possible time arguments}, which can be
either $t$ or $t-i\beta(1/2+\delta) $. The generating functional
defined in \EX spacetime by following this procedure turns out to be
the same of the one defined in the POM formalism.

\smallskip
2) One can attach the information about the region to the field
operator rather than putting it in the time argument. Thus the
identification $\phi^\I(x)\equiv \phi(x)$ and $\phi^\IId(x)\equiv
{\tilde \phi}(x)$ can be made and one obtains the formalism of TFD,
which consists of {\em two commuting field operators} and {\em a single  
time argument}.

\smallskip
Of course, the two pictures give the same physics, i.e. the same
propagator, since they are only different ``viewpoints'' of local
observers in the context of \EX spacetime.  It is an interesting
question to ask if such an equivalence can be extended for situation
which are out of thermal equilibrium. Work is in progress along this
direction.

\section*{Acknowledgements}

We thank Dr.~T.~S.~Evans for many useful discussions. M.~B.~thanks
MURST, INFN and ESF,
Y.~X.~G.~the Natural National Science Foundation of China and the Royal
Society of UK, and
F.~V.~the Swiss National Science Foundation.

\newpage

\end{document}